# Numerical computation of shape factor for pair distribution function of nanoparticles using atomistic model

Dung-Trung Tran, Gunnar Svensson and Cheuk-Wai Tai*

Department of Materials and Environmental Chemistry,
Arrhenius Laboratory, Stockholm University, Stockholm, S-106 91, Sweden
*Corresponding author: cheuk-wai.tai@mmk.su.se

**Abstract**

The atomic pair distribution function (PDF) for nano-particles has to be corrected with a shape factor, also known as form factor, in order to take size and shape effects into consideration. For most anisotropic shapes an analytical formulation of the shape factor is a challenge. In this paper, we present a new method to numerically compute the shape factor using atomistic model. This numerical method to calculate PDF for a wide range of geometrical shapes is demonstrated. In addition, a fitting formulation, which effectively converts the numerical shape factors for some selected geometrical shapes to the analytical forms, is presented.

**Introduction**

Atomic pair distribution function (PDF), as a measure of interatomic distances and coordination, is used for statistically quantitative studies of material structures (Egami & Billinge, 2002). The PDF method has been being used to quantitatively represent the atomistic structures of matters since 1930s (Tarasov & Warren, 1936; Warren *et al.*, 1936). It is particularly useful for studying gases, liquids and amorphous materials, which have no long-range periodicity. PDFs are obtained by a Fourier transformation of the structure function, which can be experimentally obtained from X-ray (Tarasov & Warren, 1936; Warren *et al.*, 1936; Warren, 1990; Egami & Billinge, 2002), neutron (Proffen *et al.*, 2003) and electron diffraction data (O'Malley *et al.*, 1998). Besides amorphous materials, crystalline materials with small particle sizes can be studied using PDFs from X-ray powder diffraction data (Warren, 1990; Proffen *et al.*, 2003). However, when using the PDF to characterize nanoscale materials one has to take into account the finite size effect (high surface-to-volume ratio) (Petkov *et al.*, 2005; Kodama *et al.*, 2006; Korsunskiy *et al.*, 2007; Gilbert, 2008; Farrow & Billinge, 2009; Lei *et al.*, 2009; Farrow *et al.*, 2010; Mullen *et al.*, 2010).

PDF is a function of the pair-interatomic distance $r$, often denoted (Egami & Billinge, 2002) as $g(r)$, describing the probability density of finding an atom pair separated by a distance $r$. For a finite size object $g(r)$ is obtained by combining the general isotropic PDF from a structure function of an infinite object with its isotropic shape factor $f(r)$ (Egami & Billinge, 2002; Kodama *et al.*, 2006; Gilbert, 2008; Lei *et al.*, 2009). The PDF for the finite object can then described as oscillating around the *shape factor* $f(r)$, giving $g(r)$:

$$g(r) = f(r) + \frac{1}{2\pi^2 r \rho_0} \int_0^\infty Q[S(Q)-1]\sin(Qr)dQ \qquad (1)$$

where $Q$ is the magnitude of the scattering vector, $S(Q)$ is the structure function, $\rho_0$ is the average number density of the material. The shape factor $f(r)$ (also called form factor) is defined (Kodama *et al.*, 2006; Gilbert, 2008; Lei *et al.*, 2009) as the ratio of the PDF $g_n(r)$ for nanoparticles to the corresponding $g_\infty(r)$ for bulk:

$$f(r) = \frac{g_n(r)}{g_\infty(r)} \qquad (2)$$

Because the probability of finding an interatomic distance longer than the maximum Feret diameter $D$ of an isolate nanoparticle is zero, the nanoparticle-PDF goes to zero at $r > D$ and hence also the corresponding shape factor $f(r)$. For an infinite large object $f(r) \equiv 1$, but for finite objects $f(r)$ is a function varying from unity to zero as $r$ increases from zero to $D$, respectively.

There is no absolute threshold in size below which finite size effects have to be taken into account. For example, when the PDF range of interest is up to 20 Å, the difference due to the finite size effect between a solid spherical particle of 100 nm in diameter and a bulk materials is around 1.5%. For a 50 nm and 10 nm particle it is ~ 3% and ~15%, respectively. In cases of porous and hollow particles, difference can be expected to be even larger. Typically, the finite size effect can be observed in nanoscale materials with sizes between around 10 nm and 500 nm, but depending on material type, PDF range and desired accuracy. The difference between a PDF calculated from an fcc bulk and the corresponding spherical nanoparticle (~4 nm in diameter) model, is shown in Figure 1.

The explicit form of the shape factor shown in Fig. (1) for a solid sphere of diameter $D$ can be written as the following (Guinier, 1963; Azaroff, 1968; Kodama *et al.*, 2006; Gilbert, 2008; Lei *et al.*, 2009) :

$$\begin{cases} f_{sphere}(r) = 1 - \frac{3}{2D}r + \frac{1}{2D^3}r^3 & \text{when } r \leq D \\ f_{sphere}(r) = 0 & \text{when } r > D \end{cases} \quad (3)$$

The shape factor $f(r)$ is also dependent on the geometrical form of the nano-particles; e.g. shape factors have been formulated for nanosheets, nanorods, nanotubes, and ellipsoids (Kodama *et al.*, 2006; Gilbert, 2008). However, these often require assumptions and may lead to complicated integrals inconvenient to evaluate. Lei et al. (2009) introduced a simulation-based method to numerically compute $f(r)$ for different shapes. In their method, a model of a desired shape, which is filled homogeneously with a huge number of points, is generated. The corresponding $f(r)$ is then equal to $g_n(r)$ as $g_\infty(r) = 1$. To achieve a desired smoothness of the computed $f(r)$, ~6.1x10$^5$ or more pairs of points per Å$^3$ were used to emulate the continuum solid medium for a tetrahedron of edge length ~2.4 nm. This large number of points needed may lead to prolonged computer time, which make it a less suitable method to correct for geometrical effects for nano-particles in reverse Monte Carlo (RMC) simulation (Thomson & Gubbins, 2000).

In this paper, we report an alternative method to compute the shape factors for various kinds of nano-particles morphologies. The advantages of our method are computational inexpensiveness and robustness. We also introduce a simple mathematical model to take into account non-elongated polyhedral shapes in the fit of the numerically computed $f(r)$. This will be demonstrated for cubic and cuboctahedral nanoparticles. How the coordination numbers of the atoms vary with size for these nanoparticles is also discussed.

**Methodology**

*Computational method*

The shape factor $f(r)$ in Eq. (2) is defined for a sphere. For anisotropic shapes the PDFs are sensitive to the relation between the orientation of the crystallographic lattice and shape of the particle. This crystallography-shape correlation has been discussed for a prolate spheroid particle model of hexagonal wurtzite ZnS, by Gilbert (2008). For elongated particles Eq. (2) is only valid for cubic and amorphous structures (effectively isotropic), according to the same author. For nanoparticles with non-elongated shapes, it is most effective to consider $f(r)$ as the angular average (Gilbert, 2008) of the corresponding $f(\vec{r})$, that is $f(r) = \overline{f(\vec{r})}$. Accordingly, the factorization described by Eq. (2) can still be applied as a smoothed-out $f(r)$ for powder samples where all possible crystallographic orientations are distributed equally with respect to the shape-related symmetry axes. If the particles have preferential crystal growth directions, then this can be accounted for in $f(r)$ by taking into account the preferential directions when constructing the appropriate atomistic models.

The direct calculation for isotropic PDFs of single-element atomistic models can be written as:

$$g(r) = \frac{1}{4\pi\rho_0 r^2 N} \sum_i \sum_j \delta(r - r_{ij}) \qquad (4)$$

where $r_{ij}$ is the pair-distance between $i^{th}$ and $j^{th}$ atoms; $N$ is the total number of atoms. It is noted that Eq. (4) can be adapted to calculate $g_n(r)$ and $g_\infty(r)$ by using appropriate pair-distance matrices $[r_{ij}]$. In order to account for the instrument functions and thermal vibrations, it is useful to convolute the $g(r)$ above with a Gaussian broadening function (Toby & Egami, 1992). This broadening causes a peak width so that the integration of $4\pi\rho_0 r^2 g(r)$ over its first peak gives the 1$^{st}$-shell coordination number $N_C^{1st}$:

$$N_C^{1st} = \int_{1^{st}\,peak} 4\pi\rho_0 r^2 g(r)dr \qquad (5)$$

In our method, the PDFs for a nanoparticle and bulk are directly computed using the corresponding atomistic model of a Bravais lattice but the construction of pair-distance matrices are different. The matrix for a nanoparticle $[r_{ij}]_n$ consists of all possible interatomic pairs in the model of $N$ atoms:

$$[r_{ij}]_n = \left|\vec{r}_i - \vec{r}_j\right|_{\substack{i=1,\dots N \\ j=1,\dots N}} \qquad (6)$$

In contrast, the matrices for bulk-PDFs $[r_{kj}]_\infty$ consists of only the pairs between $N_h$ number of the *representative central atoms* and all $N_s$ atoms in the inscribed sphere of a sufficiently large model:

$$[r_{kj}]_\infty = \left|\vec{r}_k - \vec{r}_j\right|_{\substack{k=1,\dots N_h \\ j=1,\dots N_s}} \qquad (7)$$

The representative central atoms are the non-equivalent ones in the central unit cell of the particle. With this distinct construction of the pair-distance matrices, Eq. (4) is rewritten for $g_\infty(r)$, taking into account that the range of $r$ is limited up to the radius of the inscribed sphere:

$$g_\infty(r) = \frac{1}{4\pi\rho_0 r^2 N_h} \sum_k^{N_h} \sum_j^{N_s} \delta(r - r_{kj}) \qquad (8)$$

*Analytical fitting*

The numerically computed shape factor can be converted into a parameter-based analytical form by least-squares fitting. The method is applied for non-elongated shapes, of which facets or surface symmetrically deviate from their corresponding circumscribed spheres. The analytical form of $f(r)$ is then constructed by the introduction of a multiplicative exponential factor to the formula of $f_{sphere}(r)$ in Eq. (3):

$$\begin{cases} f(r) = f_{sphere}(r,D)\gamma(r) + \varepsilon(r)\,;\ f_{sphere}(r,D)\gamma(r) \\ \gamma(r) = \exp\left\{-\alpha(D)\dfrac{r}{|r-D|}\right\} \end{cases} \qquad (9)$$

where $\gamma(r)$ can be called the *shape correction factor*; $D$ (as specified in Sec. 1) is the maximum Feret diameter of the nanoparticle, $\alpha(D)$ is a $D$-dependent fitting parameter, and $\varepsilon(r)$ is the difference between the computed $f(r)$ and the fitted $f(r)$; $f_{sphere}(r,D)\gamma(r)$. The shape correction factor satisfies the general properties of shape factors:

$$\begin{cases} \gamma(r) \to 1 & \text{when } r \to 0 \\ \gamma(r) \to 0 & \text{when } r \to D \end{cases} \quad (10)$$

For some polyhedra, the function $\alpha(D)$ has been found to conform the following simple mathematical pattern:

$$\alpha(D) = s_0 + \frac{s_1}{D} + \frac{s_2}{D^2} \quad (11)$$

where the three parameters $s_0, s_1, s_2$ can be found via fitting.

## Results

### *Features of the computed shape factor*

The modeled shape factor for a spherical fcc-structured nanoparticle model of ~ 4 nm in diameter, compared with the analytical shape factor determined by Eq. (4), is shown in Fig. 2. The modeled factor differs from the analytical one by its stepwise feature that is more visible at short distances but gradually smoothed out for increasing distances. The difference between the computed and analytical factors is given by the function $\varepsilon(r)$, which reflects the atomistic arrangement within the spherical model. If the employed model is a continuum solid (physically non-atomistic), ideally the difference between the resulting shape factor and the one obtained by Eq. (3) is negligible. However, the stepwise feature is meaningful in nanoscale particles, in which atoms are located in discrete positions. The attenuation with $r$ of $\varepsilon(r)$ in Fig. 2 shows that the computed shape factor approaches the smoothness of Eq. (3) when $r \to \infty$, equivalent to the smoothness achieved when the model approaches spatial continuum.

Spherical atomistic models of particles, with similar size but different crystal structures have been compared. The small but existing differences in crystal structure leads to slightly different shape factors. However, with increasing $r$, all the shape factors will converge to the same smoothed shape factor determined by the given size and shape. The crystal structure is less important beyond ~10 Å as demonstrated in Fig. (3) where the computed shape factors for different crystal structures (graphite, fcc, bcc, cubic-diamond and hcp) are presented for spherical models of ~5 nm in diameter. It can also be seen in Fig. (2) that the computed shape factor oscillates around the analytical curve, suggesting that least-squares fitting can be employed to smooth out the numerically computed shape factors and convert them into analytical forms depending only on size and shape of the nanoparticle. Although the crystallographic features may be potentially interesting for later studies, in this current work we only focus on the size and shape dependence of the shape factors.

*Cube-shaped nanoparticles*

Cube-shaped morphologies can be found in various nanoparticle studies. This particularly simple shape is interesting for investigations of localized surface plasmon resonances (Sherry *et al.*, 2005; Nicoletti *et al.*, 2013), magnetic properties (Salazar-Alvarez *et al.*, 2008), catalytic properties (Zhang et al., 2008), and nanocrystal growth (Liao *et al.*, 2014). One model to approximate the shape factor for a cube of edge length $L$ has been reported by Korsunskiy *et al.*(2007):

$$f_{cube}^{approximated}(r,L) = 1 - \frac{3}{2L}r + \frac{2}{\pi L^2}r^2 - \frac{1}{4\pi L^3}r^3 \qquad (12)$$

However, Eq. (12) is valid only for $r$ distances smaller than the diameter $L$ of the inscribed sphere of the cube. For this $r$ range, the shape factor for the cube was claimed not to deviate significantly from that for a sphere of a comparable size (Korsunskiy *et al.*, 2007). Besides the approximation given by Eq. (12), a full derivation of $f(r)$ in closed form for cube-shaped nanoparticles has previously not been reported in literature to our knowledge. This is mainly due to that edge and vertex singularities are difficult to handle with closed form formulations (Lei *et al.*, 2009).

An alternative approach is to use atomistic models describe the shape function. This is here illustrated for cube-shaped fcc nanoparticles, where the size-dependent shape factor has been computed for particles of different maximum $D$. The computed shape factor numerically calculated for a cube-shaped model particle built from 3430 atoms having the maximum Feret diameter $D \approx 5.33$ nm and the edge length $L = D/\sqrt{3}$, is shown in Fig. (4). It is interesting to note that the crystallographic feature of this computed shape factor does differs from the stepwise feature observed for the spherical fcc models [see Fig. (2) and Fig.

(3)] by exhibiting complicated sharp ripples. This is due to the anisotropy of the cubic shape, as mentioned above in Sec. 2.1 and previously by Gilbert (2008). These sharp ripples, which are caused by the relative orientations between the crystallographic axes and surfaces, will obviously cause a biased scaling for PDFs in Eq. (2). Fortunately, a least-squares fitting can be employed to smoothen the computed shape factor. This excludes the crystallographic features and recover the size and shape-dependence of the shape factor curve.

The fitting curve for the computed shape factor using Eq. (9), is presented as the blue-solid curve in Fig. (4). The corresponding root-mean-square (RMS) error for this fitting is ~1.3%, mainly caused by the crystallographic ripples, as illustrated by the $\varepsilon(r)$ function (difference between computed and analytical factors). For comparison, the previously reported approximated shape factor [see Eq. (12)] for the cube-shaped model and the shape factor for a sphere [see Eq. (3)] having the same volume of the cube are also shown in Fig. (4) by green-dash and black-dot curves, respectively. It is seen that the approximation specified by Eq. (12) (Korsunskiy *et al.*, 2007) significantly differ from the numerical model presented here. This is understandable as Eq. (12) was derived using the same approach with the one used for spheres (Korsunskiy *et al.*, 2007) while the morphology of cubes is highly different from that of spheres. The volume of a cube is only ~36.8% of the volume of its circumscribed sphere.

It is very often that the corners of real cube-shaped nanoparticles are truncated. The effect of this was investigated by computing the shape factors for a complete cube with single-atom vertexes and incomplete ones with missing vertexes, both with fcc type structures. For illustration, Figures 5a and b shows the [111] projections of a 3430-atom complete cube model and a 4630-atom incomplete cube model with eight missing vertexes, respectively. For the incomplete cube, the vertex-truncated areas are 3-atom triangles. The shape factors for the complete and incomplete cubes are found to be significantly different from each other. Figure 6a shows the size-dependent $\alpha(D)$ parameters computed and fitted for various sizes of complete and incomplete cobalt-fcc cube models. The size-dependence curves for $\alpha(D)$ are plotted by fitting using Eq. (11). The significant differences between the $\alpha(D)$ curves for the complete and incomplete cubes suggest that the PDFs for cube-shaped nanoparticles are quite sensitive to their surface construction even at the atomic level. The averaged first coordination numbers [Fig. (6b)], however, are less sensitive to the atomic arrangement at the surface. The size-dependence of these coordination numbers was plotted by fitting:

$$N_C^{1st}(D) = 12 + \frac{c_1}{D} + \frac{c_2}{D^2} \qquad (13)$$

where $c_1$ and $c_2$ are the fitting parameters. It is noted that Eq. (13) is applied for fcc and hcp structures having the bulk coordination number of 12. The fitting parameters corresponding to Fig. (6) are presented in Table (1).

*Cuboctahedral nanoparticles*

The cuboctahedral morphology which can be described as special case of a truncated cube is very frequently found for nanoparticles (e.g. Rodriguez *et al.*, 1996; Montejano-Carrizales *et al.*, 1997; Pauwels *et al.*, 2001; Frenkel, 2007; Stephanidis *et al.*, 2007; Farrow *et al.*, 2010; Tran *et al.*, 2011). Atomistic models of cuboctahedra shaped with triangular (111) and square (100) facets can be built using a fcc structure. The number of atoms needed for a complete cuboctahedron is called the magic number (Frenkel, 2007). The relation between size, shape factors and coordination numbers have been studied for fcc cuboctahedra of different sizes corresponding to the magic numbers 309, 923, 2057, 3871, 6525, 10179, and 14993. Figure 7 shows the computed shape factor for a cuboctahedron built from 3871 atoms ($D \approx 5.76$ nm) and the corresponding fitting using Eq. (9). The fitting error is 0.68%, significantly smaller than that (1.3%) in the model of the cube having a comparable size ($D \approx 5.33$ nm) presented in Sec. 2.1. This is not surprising as a cuboctahedron is less anisotropic than a cube, so the ripples of crystallography-shape correlation are weaker in the cases of cuboctahedra and the volume of a cuboctahedron is ~56.3% of its circumscribed sphere's volume (compared with ~36.8% for a cube).

The size-dependent $\alpha(D)$ parameters computed and fitted for the above-mentioned cuboctahedra, is shown in Figure 8a. The size dependence of the first coordination numbers of cuboctahedra has been analytically investigated before (Montejano-Carrizales *et al.*, 1997). The following equation presents an exact formulation:

$$N_{C(cubo)}^{1st} = \frac{24\Gamma(5\Gamma^2 + 3\Gamma + 1)}{10\Gamma^3 + 15\Gamma^2 + 11\Gamma + 3} \qquad (14)$$

where $\Gamma = D/(2r_{min})$ is the cluster order, with $r_{min}$ as the minimum interatomic distance. An approximation (Frenkel, 2007) has also been presented, applied only for large ($D \gg r_{min}$) cuboctahedra:

$$N^{1st}_{C(cubo)} \approx 12 - \frac{9}{\Gamma} + \frac{3}{4}\frac{1}{\Gamma^3} \qquad (15)$$

Fig. (8b) shows the computed coordination numbers for the cuboctahedra and the corresponding fitting using Eq. (13) alongside the results obtained by Eq. (14) and (15) for comparison. The computed values from the model presented here are found to match exactly with Eq. (14) while Eq. (15), which approximates large cuboctahedra as spheres, seems to be less applicable in the presented size range. This confirms that our computations of coordination numbers using the PDF approach [see Eq. (5)] are of very high accuracy. The fitting parameters corresponding to Fig. (8) are presented in Table (1).

**Discussion**

Our method for computation of shape factors is robust because only a non-simulated routine is required. The routine is building atomistic models with translational symmetry for bulks and the desired nanoparticle shapes, then computing the bulk and corresponding nanoparticle PDFs. The minimum requirement for the size of the bulk model depends on how large the nanoparticles are and how long the $r$-ranges that are needed for the shape factors. For example, to compute the full-range (up to $D$) shape factor for a fcc cuboctahedral model having $D \approx 9.2$ nm, a spherical bulk model having the radius larger than ~9.2 nm, corresponding to at least $\sim 2 \times 10^5$ atoms, is needed. Because the hierarchical number for the fcc structure is 1 (for fcc structure, all atoms are equal in position), the bulk-matrix, according to Eq. (7), consists of only $\sim 2 \times 10^5$ pair distances. This number is relatively insignificant when compared with the nanoparticle-matrix [Eq. (6)] for the cuboctahedral model that consists of around $2 \times 10^8$ pair-distances. The total computation time will strongly depend on the number ($N$) of atoms consisted in the nanoparticle model, as the number of pair-distances is equal to $N^2$. To put the computation cost in the context of comparison, it is noted that $10^9$ pairs of points were needed to compute the shape factor for a tetrahedron of edge length ~2.4 nm using the approach of continuum solid (Lei *et al.*, 2009).

The fitting using Eq. (9) can be applied only for non-elongated polyhedra those have no one-dimensional size extension. Apart from elongation problems, the fitting accuracy is sensitive to the anisotropy of polyhedra. Presented in Table (1), the deviations of the parameters for cubotahedra are generally much smaller than for cubes. This is due to the ripples of crystallography-shape correlation as discussed above in Sec. 2.1 and 3.2. These ripples are expected to be stronger for shapes of more anisotropy, particularly in the cases of non-cubic structures, causing larger deviations for the fitting parameters. In the cases of elongated shapes, although using Eq. (9) is not recommended, our computation method is still applicable for cubic structures and the smoothing out the computed shape factor may still be possible by using appropriately polynomial fitting.

**Summary**

A distinct method to numerically compute shape factors, which are the characteristic of the finite size effect in PDFs, is presented. It can be applied to the atomistic models with various shapes. The resulting computed shape factors contain crystallographic features, which can correlate with the anisotropy of the particle shapes. In order to offer a convenient evaluation of shape factors, an analytical fitting model was introduced for non-elongated shapes. Computation and fitting have been demonstrated for cube-shaped and cuboctahedral nanoparticle models.

Acknowledgement

The Knut and Alice Wallenberg (KAW) Foundation is acknowledged by the authors for providing the financial support under the project 3DEM-NATUR.

**Table 1**. Fitting parameters listed for complete cubes, 4-vertex-truncated cubes, 8-vertex-truncated cubes, and cuboctahedra.

| Shapes | $\alpha(D) = s_0 + \dfrac{s_1}{D} + \dfrac{s_2}{D^2}$ | | | $N_C^{1st}(D) = 12 + \dfrac{c_1}{D} + \dfrac{c_2}{D^2}$ | |
|---|---|---|---|---|---|
| | $s_0$ | $s_1 (Å)$ | $s_2 (Å^2)$ | $c_1 (Å)$ | $c_2 (Å^2)$ |
| complete cube | 0.937(4) | -1.03(3) | 0.18(4) | -6.97(6) | 2.26(7) |
| | ±0.0009 | ±0.007 | ±0.009 | ±0.015 | ±0.032 |
| 4-vertex-truncated cube | 0.916(8) | -1.24(9) | 0.31(2) | -6.88(3) | 2.51(3) |
| | ±0.0029 | ±0.023 | ±0.037 | ±0.032 | ±0.06 |
| 8-vertex-truncated cube | 0.933(7) | -1.61(4) | 0.65(7) | -6.67(4) | 2.37(1) |
| | ±0.0009 | ±0.007 | ±0.012 | ±0.066 | ±0.078 |
| cuboctahedron | 0.452(1) | -0.79(3) | 0.18(6) | -6.20(7) | 1.73(8) |
| | ±0.0001 | ±0.001 | ±0.002 | ±0.019 | ±0.055 |

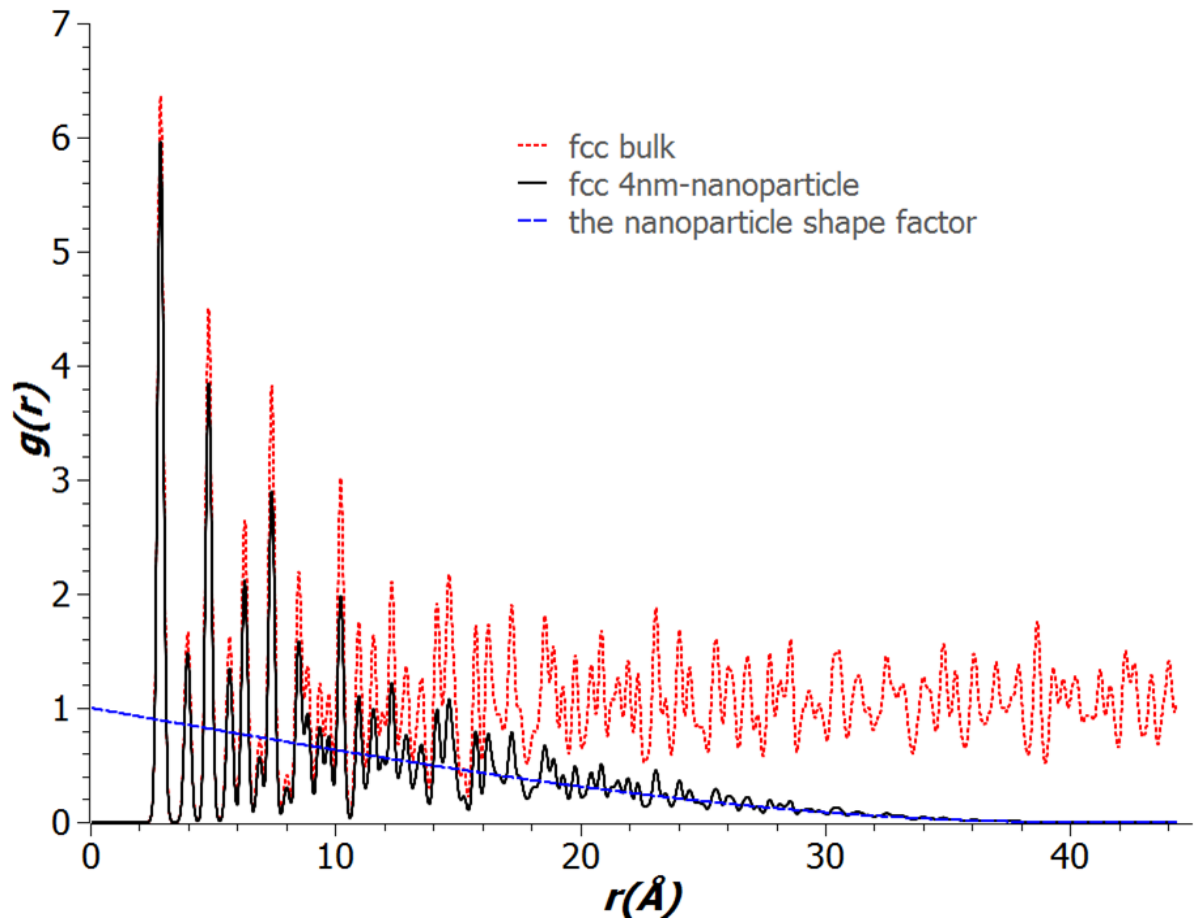

**Figure 1**. PDFs calculated for atomistic models of fcc structure: bulk (red dot), a spherical nanoparticle of ~ 4nm in diameter (black solid); the corresponding shape factor is shown by the blue dash line; the black arrow indicates where the nanoparticle PDF and the shape factor start to vanish.

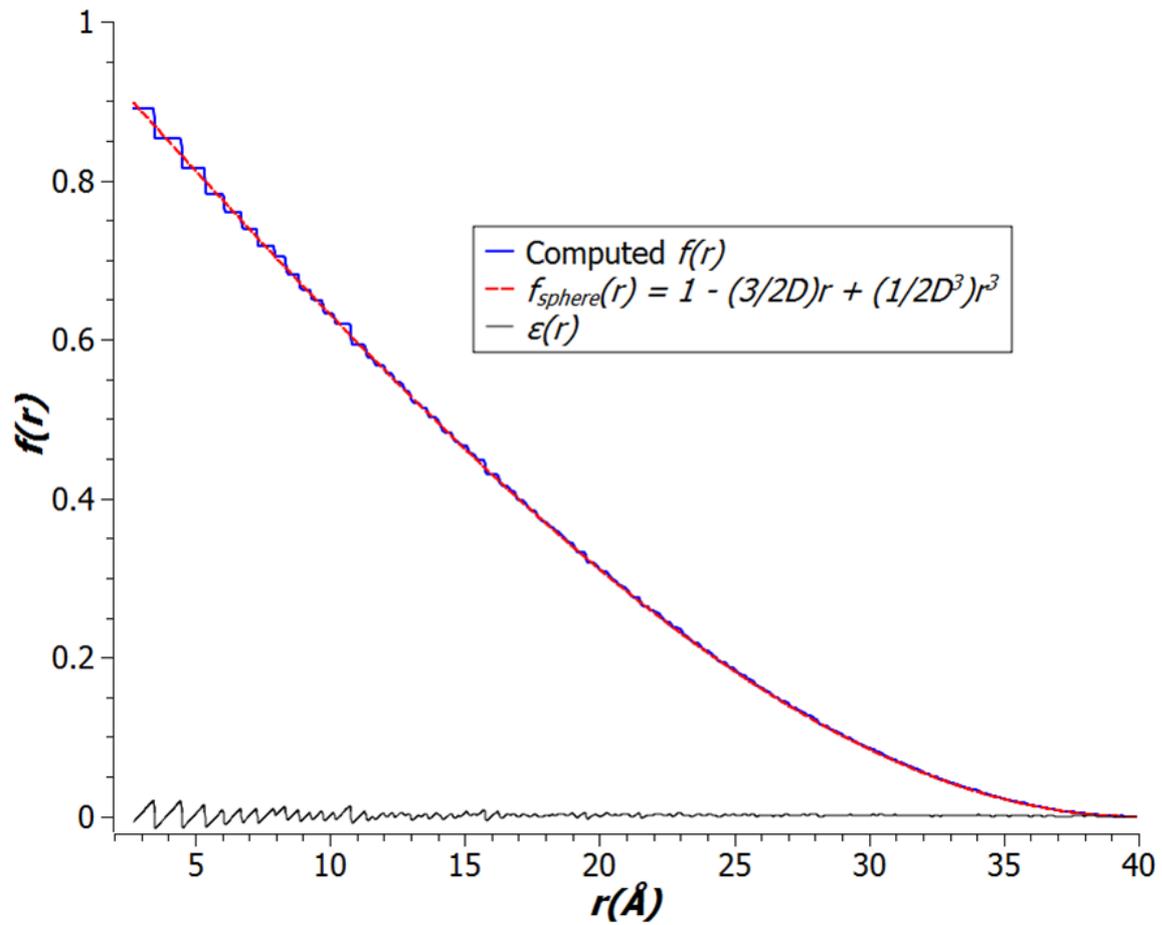

**Figure 2**. The computed shape factor (blue-solid) for a ~4nm-diameter spherical nanoparticle having an fcc crystalline structure compared with Eq. (3) (red-dash) for a sphere of the same diameter; the difference $\varepsilon(r)$ between the two factors is shown by the black-solid ripples.

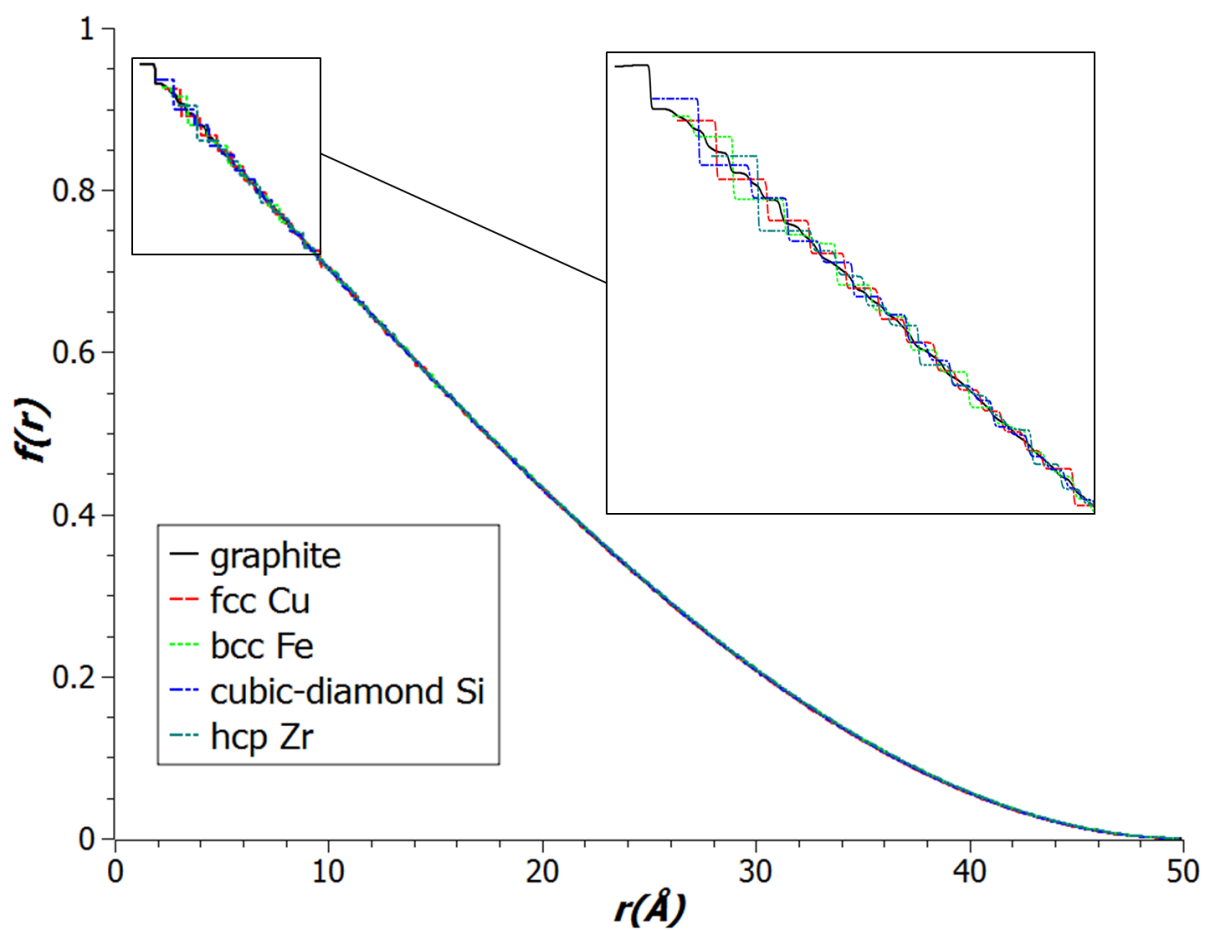

**Figure 3**. Computed shape factors for ~5 nm-diameter spherical models having different crystalline structures: graphite, fcc copper, bcc iron, diamond silicon, and hcp zirconium; the inset shows a magnified region ( $r <\sim 10 Å$ ) for the crystallographic features to be more discernable.

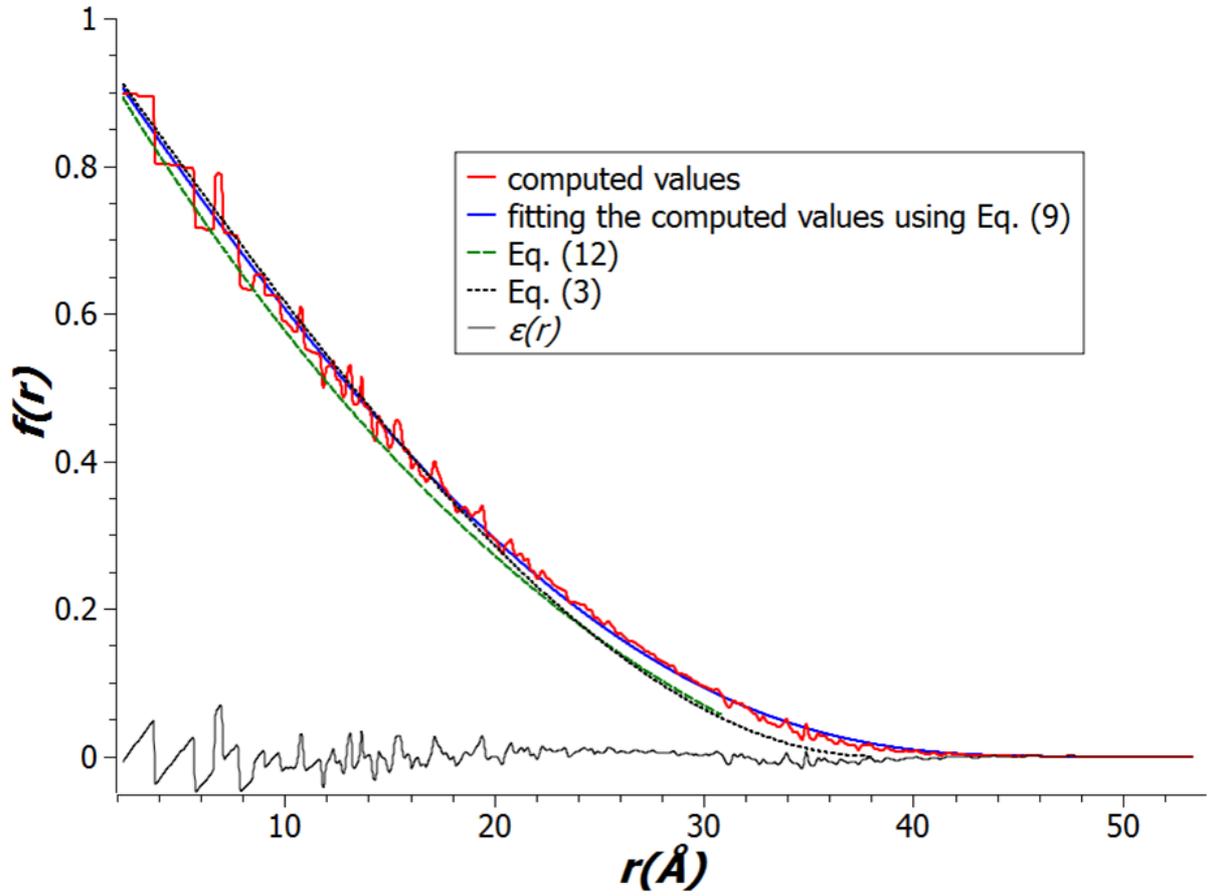

**Figure 4**. Computed shape factor (red solid) for a cube-shaped fcc model consisting of 3430 atoms and having a maximum Feret diameter of $D \approx 5.33$ nm; the blue solid curve presents the fitting of the computed factor using Eq. (9); the difference $\varepsilon(r)$ between the fitting and the computation is presented by the bottom black solid line; the shape factor previously approximated by Eq. (12) is presented by the green dash curve up to the edge length $L = D/\sqrt{3}$; the shape factor calculated using Eq. (3) for a sphere having the same volume of the cube is presented by the black-dot curve.

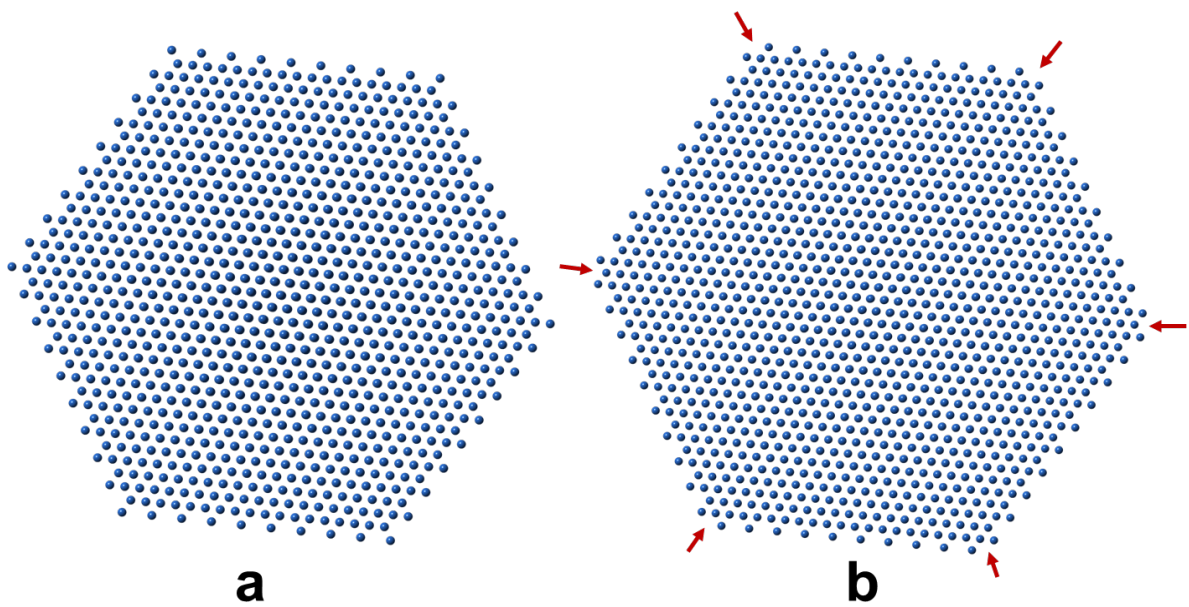

**Figure 5**. Projections along the [111] direction of a 3430-atom complete cube model (a) and a 4630-atom incomplete cube with eight truncated vertexes (b); six missing vertexes visible in the current projection are marked by arrows.

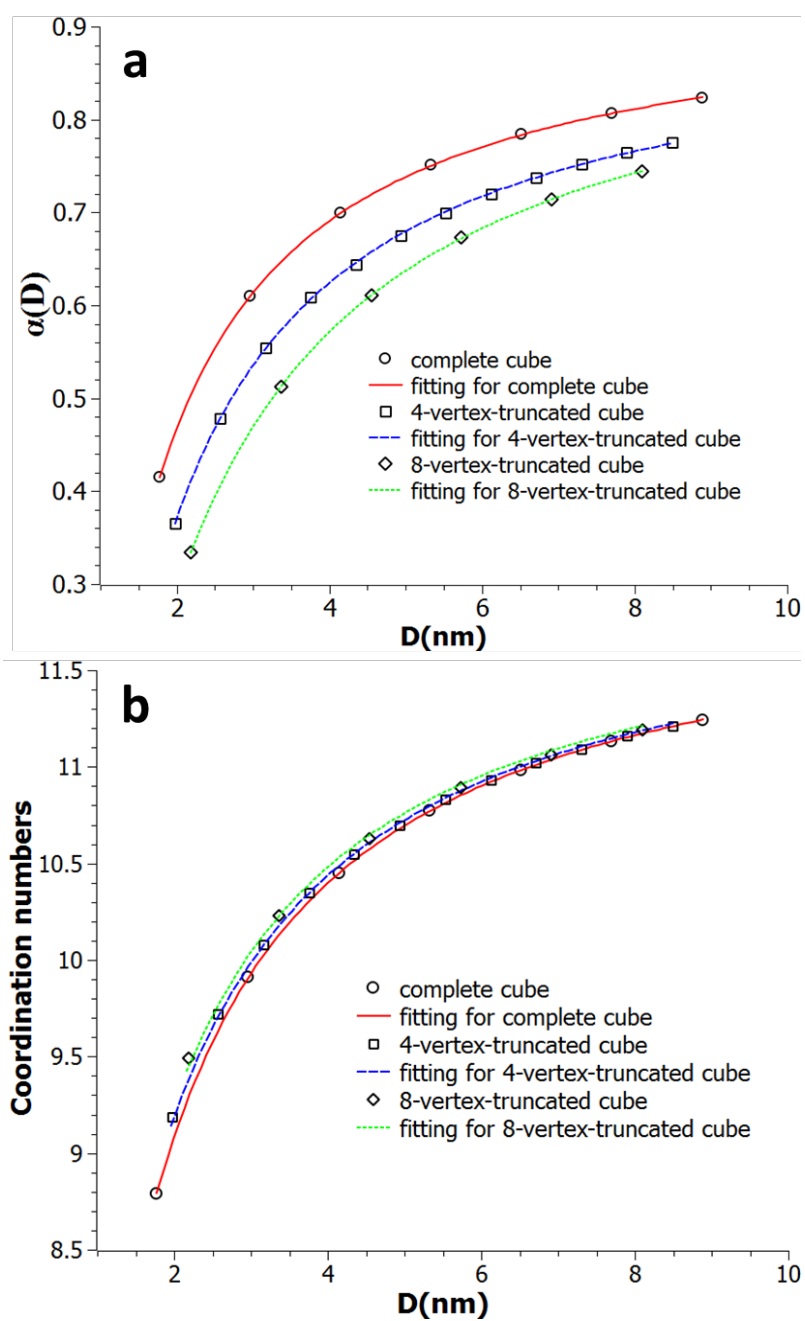

**Figure 6**. (a) Size-dependent $\alpha(D)$ and (b) coordination numbers computed and fitted for complete and incomplete cube models of various sizes. The values for perfect cubes (circle and red solid), incomplete cubes with four vertexes truncated (square and blue dash), and incomplete cubes with eight vertexes truncated (rhombus and green dot) are plotted and fitted.

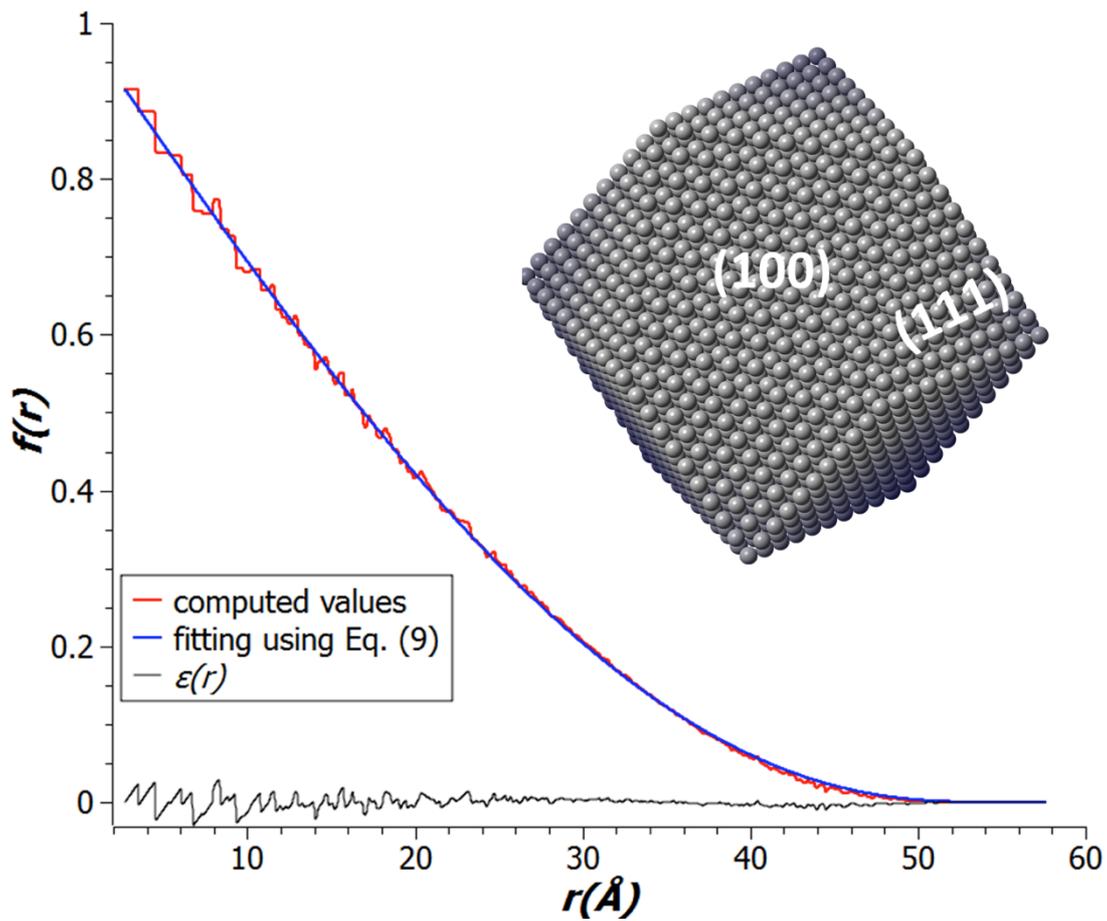

**Figure 7**. Computed shape factor (red-solid) for a fcc cuboctahedron consisting of 3871 atoms and the corresponding fitting curve (blue-solid). Their difference is shown as black solid line. The corresponding nanoparticle model is shown in the inset.

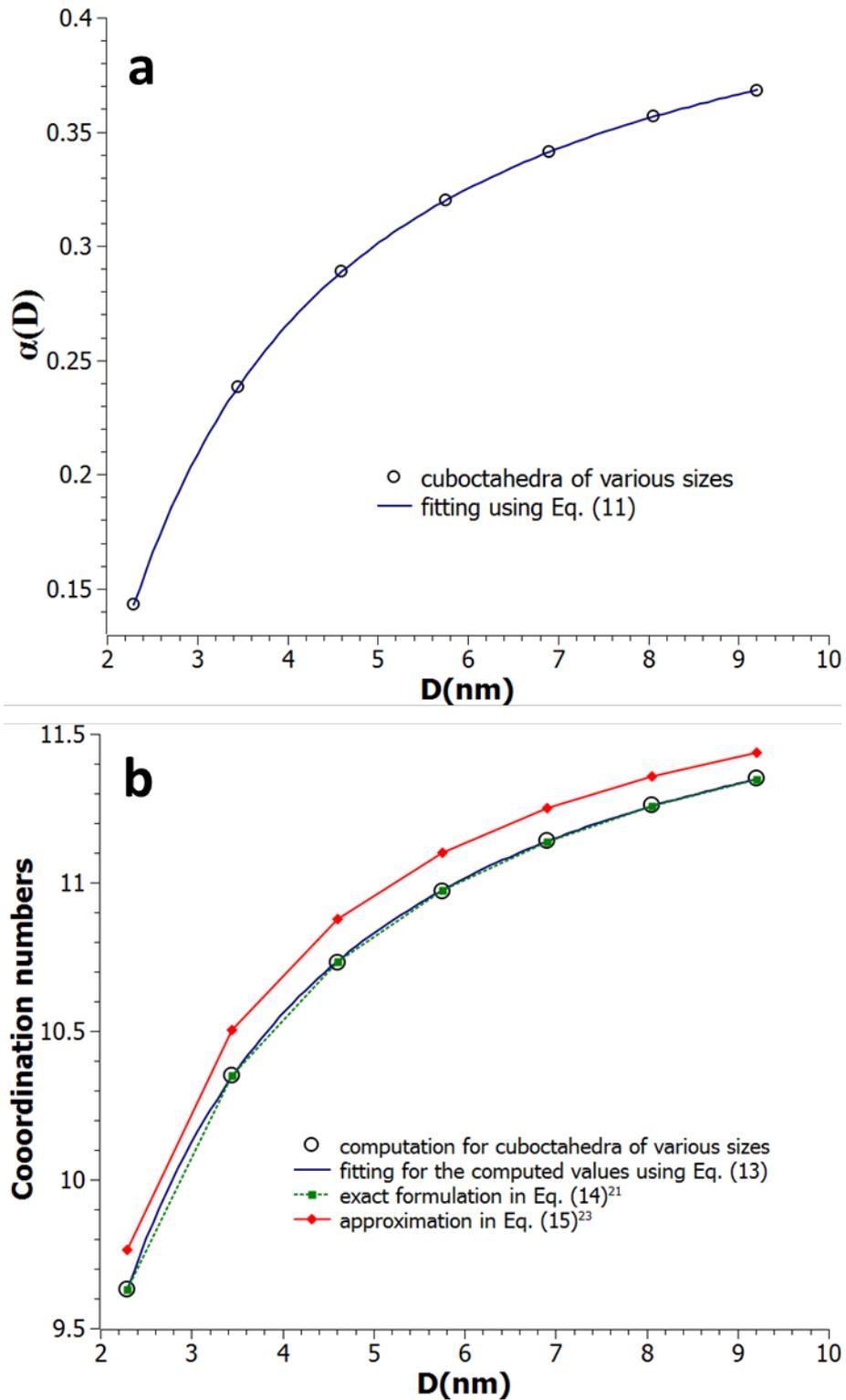

**Figure 8**. (a) $\alpha(D)$ for cuboctahedra of various sizes (black circle) and the corresponding fitting (blue-solid); (b) the first coordination numbers shown with computed values (black circle), fitting for the computed values (blue solid curve), Eq. (14) calculation (green square), and Eq. (15) calculation (red solid rhombus).